\documentclass{article}
\usepackage[onecolumn]{emulateapj}
  \newcommand {\Ha}     {H$\alpha$}
  \newcommand {\HH}     {H$_2$}
  \newcommand {\HI}     {\ion{H}{1}}
  \newcommand {\NaI}    {\ion{Na}{1}}
  \newcommand {\OI}     {\ion{O}{1}}
  \newcommand {\NII}    {\ion{N}{2}}   
  \newcommand {\FeII}   {\ion{Fe}{2}}  
  \newcommand {\CI}     {\ion{C}{1}}   

  \newcommand {\kms}      {km~s$^{-1}$}

\begin{document}

\title{The ISM Interactions of a Runaway LBV Nebula in the LMC}

\author{Charles W. Danforth\altaffilmark{1}, You-Hua Chu\altaffilmark{2}} 
\altaffiltext{1}{Department of Physics and Astronomy, The Johns Hopkins University, 3400 N. Charles Street,  Baltimore, MD 21218; danforth@pha.jhu.edu} 
\altaffiltext{2}{Astronomy Department, University of Illinois, 1002 W. Green Street, Urbana, IL 61801; chu@astro.uiuc.edu} 


\begin{abstract} 
New observations of the Magellanic Cloud Luminous Blue Variable candidate S119 (HD~269687) show the relationship of the star to its environs.  Echelle spectroscopy and high-resolution HST imagery reveal an expanding bubble centered on the star.  This bubble appears in both \Ha\ and [\NII] and is noticeably brighter on the near (blue-shifted) side.  The systemic velocity of both the expanding bubble and the star itself (as seen by the very broad \Ha\ emission feature in the stellar spectrum) is V$_{\rm hel} \sim$160 \kms\ whereas the velocity of the superposed LMC ISM is 250--300 \kms.  ISM absorption features seen in FUSE spectra reveal components at both stellar and LMC velocities.  Thus we conclude that S119 is located within the LMC ISM and that the bubble is interacting strongly with the ISM in a bow shock.
\end{abstract}


\section{Introduction}
Small ring nebulae have been detected around luminous blue variables (LBVs) in the Milky Way and the Large Magellanic Cloud (LMC) (Nota et al. 1995).  Spectroscopic observations of such nebulae show enrichment of nitrogen and depletion of oxygen, indicating that these nebulae contain processed stellar material ejected by the LBVs (e.g., Smith et al. 1998).  A large range of expansion velocities are observed in the dozen or so known LBV nebulae: the $\eta$ Car nebula has the largest expansion velocity, $\sim$600 \kms; a few nebulae have expansion velocities $\ge$ 50 \kms, e.g., P Cyg, HR Car, and AG Car; and the slowest expansion velocities are 20-30 \kms, e.g., R127 and S119 (Nota et al. 1995).  These diverse expansion dynamics indicate that either these LBV nebulae are at different evolutionary stages or they have different formation mechanisms.

Two LBVs projected against the LMC, R127 and S119, are surrounded by ring nebulae with similar angular sizes and expansion velocities (Nota et al. 1995); however, they have very different radial velocities.  R127 has a radial velocity consistent with the systemic velocity of the LMC, V$_{\rm hel}$ = 250--300 \kms, suggesting it is a member of the LMC (Walborn 1982).  S119, on the other hand, has a radial velocity that is blue-shifted $\sim$100 \kms\ from the LMC systemic velocity.  Furthermore, the interstellar \NaI\ absorption in the spectrum of S119 does not show an LMC component (Nota et al. 1994).  Thus the membership of S119 in the LMC cannot be automatically assumed.

Whether S119 was formed in the LMC or the Milky Way, its velocity implies that it has been ejected by its host galaxy at a high speed.  The evolutionary status of S119 is thus very interesting.  We have obtained a high-dispersion, long-slit echelle observation of the nebula around S119 and high-dispersion far ultraviolet spectrum of the star S119.  These spectra are used to examine the dynamics of the nebula, study the interaction between the LBV nebula and the ambient interstellar medium (ISM), and determine whether S119 belongs to the Galaxy or the LMC.

\section{Observations}
On 2000 December 9, we obtained a long-slit, high-resolution spectrum of S119 (aka HD~269687, aka Sk$-$69$^\circ$175) using the echelle spectrograph on the CTIO 4-meter telescope.  The principal aim of the observing program was to investigate the structure of the Magellanic Cloud ISM near approximately 70 hot stars observed by the Far Ultraviolet Spectroscopic Explorer (FUSE).  The data presented here was a 900 second exposure with a 1\farcs64$\times$4\arcmin\ slit oriented east-west.  A post-slit \Ha\ filter ($\lambda_c$=6563 \AA; $\Delta\lambda$=75 \AA) restricted our spectral coverage to 6506--6654 \AA, a single echelle order which contains \Ha\ and [\NII]$\lambda\lambda$6548.05, 6583.45 lines.  The spectrograph resolution was approximately 12 \kms$\times$2\farcs3.  The data were reduced using standard techniques within IRAF and are presented in Figure~1.

To supplement our high resolution optical spectrum, we have inspected the FUSE observation of S119.  Technical specifications of the FUSE spacecraft are described in Moos et al. (2000) and Sahnow et al. (2000).  The single 3780~s FUSE observation (P11748) was taken on 2000 October 2 and covers the wavelength range 905-1187 \AA\ at a resolution comparable to our echelle data ($\sim$15 \kms).  The FUSE spectral range is unusually rich in ionic, atomic and molecular transitions.  We will use the FUSE data to probe gas seen in absorption along the line of sight to S119.  Of particular interest are the \OI\ $\lambda$1039.23 which probes total \HI\ column and \FeII\ $\lambda$1144.94 which probes ionized material similar to that shown by \Ha\ and [\NII].  Additionally, we examine \CI\ $\lambda$945.191 and several \HH\ lines which probe cold gas.

Finally, we consider archival HST imagery of S119 in our analysis.  Four WFPC2 exposures from 1998 March observed with the \Ha\ (F656N) filter were coadded for a total exposure length of 2000 seconds.  As the star was centered in the Planetary Camera for this observation, we see the circumstellar nebula at unprecedented resolution.  This image is presented in Figure~2.

\section{Discussion}
\subsection{Expanding Shell}
The most obvious feature of Figure~1 is the expanding shell seen in both \Ha\ and [\NII].  This shell is easily spatially resolved along the slit and shows a diameter of 7\farcs5$\pm$0\farcs5 or $\sim$2 pc at the distance of the LMC.  In velocity space, the shell shows well-resolved expansion with two sides separated by about 50 \kms\ for an expansion velocity of 25 \kms.  This expansion velocity is on the low side for LBV nebulae but is comparable to that of R127.  The red and blue-shifted components of the shell in each line show a width only slightly higher than the resolution of our spectrograph ($\sim$15 \kms).  

Measuring the east-west diameter of the bubble in Figure~2, we find 7\farcs2$\pm$0.5\arcsec, though the western edge of the nebulosity is somewhat hard to define.  In the north-south direction, we get a slightly longer diameter of 7\farcs8$\pm$0\farcs5 (not including the short tail to the southwest).  For comparison, the diameter and expansion velocity reported by Nota et al.\ (1994), measured from ground-based data, are 8\arcsec$\times$9\arcsec\ and 25 \kms.

\subsection{Where is S119?}
By comparing the emission features seen in the optical spectrum with the absorption features in the FUSE spectrum, we can investigate the three-dimensional structure of S119 and its surroundings.  By looking at low-ionization gas (\OI, \FeII, etc.) we should see in absorption much of the same material seen in emission in \Ha.  

The upper panel of Figure~3 shows the observed velocity structure in \Ha\ and [\NII] extracted from our echelle data.  The solid curves show the stellar spectrum in \Ha\ and [\NII] and the dashed curve shows the averaged spectrum of an adjacent few arcseconds of ISM in \Ha.  We see a pair of emission peaks at 140 and 190 \kms\ corresponding to the expanding shell.  In the background spectrum, we see a broad band of LMC \Ha\ emission between 220 and 290 \kms. 

The lower panel of Figure~3 shows the normalized absorption profiles of \OI\ $\lambda$1039 and \FeII\ $\lambda$1145 from the FUSE data.  At least five distinct velocity components are detected with good agreement between the two species.  A strong line at 20 \kms\ arises in local Galactic material.  This is consistent with the absorption components at 18 \kms\ seen by \cite{RothBlades97} in \ion{Zn}{2} and \ion{Cr}{2}, two ions with very similar ionization potentials to \FeII.  At 60 \kms\ we see what is most likely an intermediate-velocity cloud (probably located in the Galactic halo).  

There are three higher velocity absorption components seen at 135, 215 and 290 \kms.  The strong feature at 135 \kms\ may be associated with the near (blue-shifted) side of the LBV nebula seen in emission at 140 \kms.  However, this feature may also arise in high velocity cloud material from the Galactic halo which is often seen at $\sim$125 \kms\ toward the LMC (Richter et al. 1999).  We do not expect, nor do we detect any absorption features corresponding to the receding side of the bubble at 190 \kms.

The 215 and 290 \kms\ absorption components represent LMC ISM absorption and are typical of LMC sightlines (Danforth et al. in prep).  Roth \& Blades see an absorption component at 283 \kms\ which is almost certainly the same as our 290 \kms\ component.  A reasonably good correlation is seen between these absorption features and the \Ha\ emission at 220--290 \kms;  they may trace the same material.  Since LMC gas is seen in absorption, S119 must be behind some portion of the LMC ISM.  

At first glance, this appears to be in contradiction to the findings of Nota et al. (1994) who detected no absorption in \NaI\ at LMC velocities.  Na is far less abundant than either O or Fe in the solar neighborhood (presumably in the metal-poor LMC as well) and thus a non-detection of \NaI\ does not rule out a detectable quantity of \OI\ or \FeII.  Furthermore, \NaI\ has an ionization potential of only 5.1 eV and is found in the cold, neutral clouds, while \OI\ is ionized at 13.6 eV and traces the total \HI\ column and can exist in conciderably warmer conditions.  

Better tracers of cold gas are \CI\ which has an ionization potential of 11.26 eV and molecular hydrogen which is dissociated at 4.48 eV.  The The FUSE data do not show significant \HH\ or \CI\ $\lambda$945.191 absorption at LMC velocities, suggesting that there is a very low column depth of cold neutral gas in our line of sight.  The ISM along the sightline toward S119 must be dominated by ionized gas at warm to high temperatures, which can be detected in \OI\ but not \NaI\ or other cold gas tracers.

\subsection{ISM Interactions}
The HST image reveals that the northeastern side of S119's nebula  is brighter and has a sharper outer edge than the southwestern side.  The image also shows that S119 is slightly displaced from the nebular center toward the northeast.  These characteristics are typically seen in planetary nebulae moving through the ISM (Tweedy \& Kwitter 1996). The echelle image reveals that the near (blue-shifted) side of the bubble is brighter than the far (red-shifted) side by a factor of 3.3 in both \Ha\ and [\NII] lines.  While we may simply be seeing an inhomogeneity in the nebular density or internal extinction in the nebula, we have shown that S119 lies behind a significant amount of LMC ISM and is moving through it at $>$100 \kms.  These points seem to indicate that the bubble is interacting with the ISM on the near side in a bow shock.  Taking the sharp, bright northeastern shell rim as the leading edge, we can conclude that the star is moving away from the LMC (toward us) and to the northeast at a sizable velocity.

The rapid movement of S119 and its nebula through the LMC ISM is further supported by the detection of the `wake' of the bow shock in the echelle observation.  Faintly seen in Figure~1 both in \Ha\ and the [\NII]~$\lambda$6584 lines is a `tail' of emission stretching from the western tip of the expansion feature back toward the bulk of emission representing the LMC gas.  This tail is brighter toward the bubble velocity than toward the LMC ISM velocity, indicating that it contains the nebular material that has been decelerated by the LMC ISM. 

\section{Conclusions}
S119 has long been known to be an LBV candidate star with an expanding nebular ring.  In this letter we have confirmed the expansion velocity of 25 \kms\ and refined the physical scales of the bubble to 7\farcs2$\times$7\farcs8 or 1.8$\times$2.0 pc at the distance of the LMC.  

Given the anomalous stellar velocity of S119, v$_{\odot}\sim$160 \kms, membership in the LMC (or in the Galaxy) was in debate.  The presence of low-ionization absorption at LMC velocities against the stellar continuum indicates that S119 resides within the LMC.  Though Nota et al.\ (1994) did not see \NaI\ absorption at LMC velocities, a similar dearth of \CI\ and \HH\ suggest that the S119 sightline contains very little cold neutral gas.  The star's anomalously high velocity might be caused by the supernova explosion of a close companion star (the ``sling-shot'' effect; Blaauw 1961).  It would be extremely interesting to search for a compact companion of S119.

The relative brightening on the near side of the expanding nebular shell as well as the velocity disparity between S119 and the LMC gas leads us to postulate a bowshock interaction.  The appearance of a faint `tail' on the western shell edge (Figure 1) and the relative brightening and sharp edge of the northeastern edge of the nebular shell (Figure 2) suggest a general northeastward motion as well.

\section{Acknowledgements}
The authors wish to acknowledge Alex Fullerton and FUSE Hot Star team for use of their proprietary FUSE data.  Discussions with Chris Howk, Antonella Nota and Nichole King were particularly helpful as well.  This work is based in part on data obtained for the Guaranteed Time Team by the NASA-CNES-CSA FUSE mission operated by the Johns Hopkins University.  Generous travel support was provided by CTIO.  Financial support has been provided by NASA contract NAS5-32985.

\begin{figure}
\epsscale{1.0}\plotone{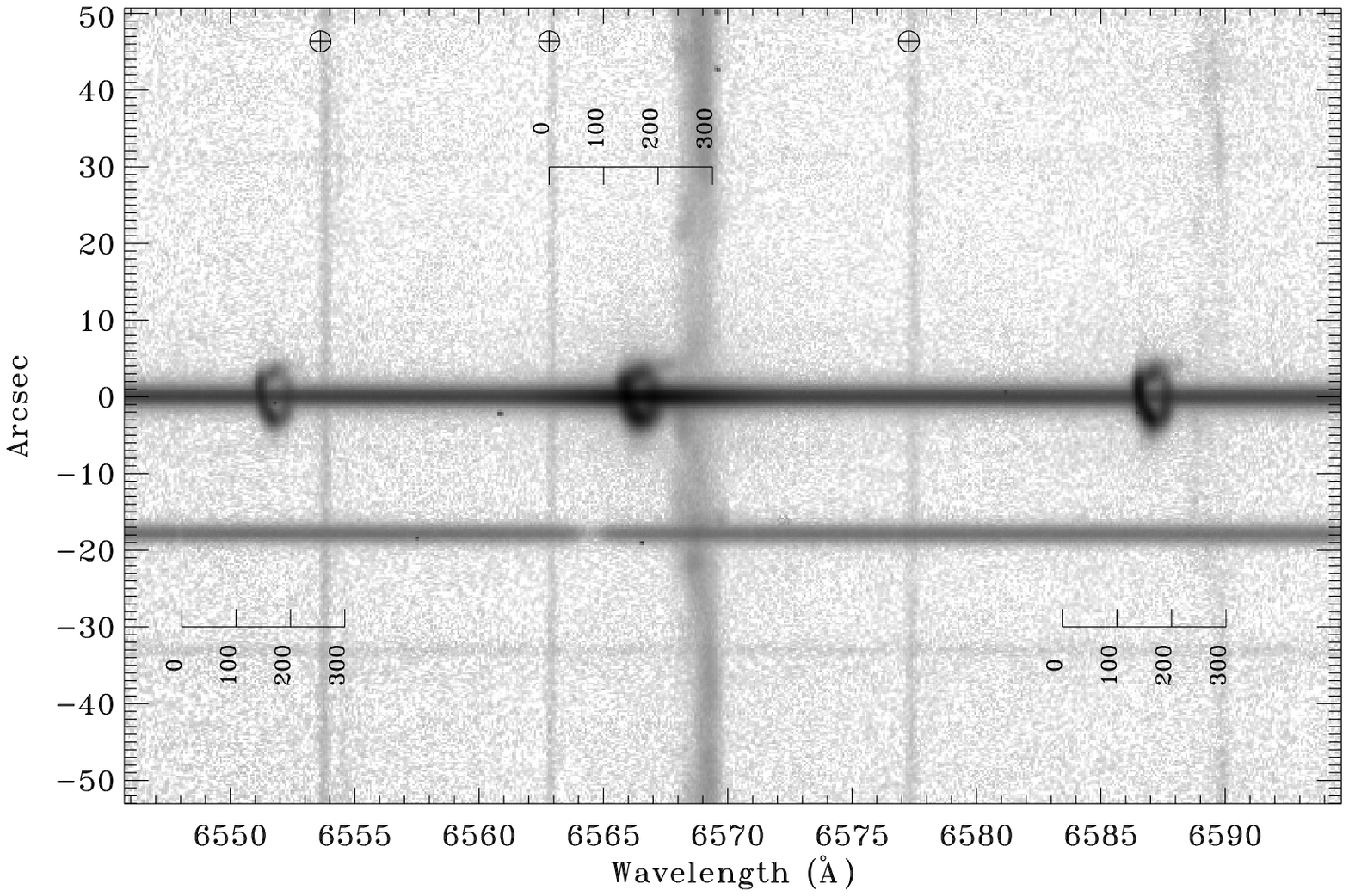}
\caption{An echelle spectrum of S119.  The slit was oriented east-west and west is at the image top.  Internal velocity axes have been plotted for each emission line in units of \kms.  The image is displayed logarithmically.  A bright expansion feature can be seen in \Ha\ and both [\NII] lines.  The LMC ISM can also be faintly seen at 240--300 \kms\ in \Ha\ and faintly in [\NII]$\lambda$6583.  The narrow line at 260 \kms\ in the [\NII] $\lambda$6548 line is terrestrial.  The continuum source at $\sim18''$ east of S119 is a foreground Galactic star.}
\end{figure}

\begin{figure}
\epsscale{.5}\plotone{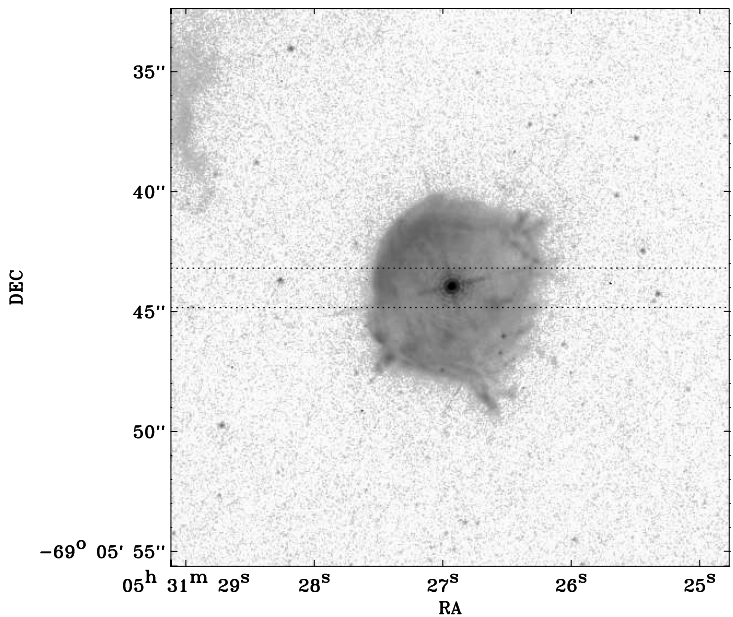}
\caption{Planetary Camera image of S119 captured by HST in \Ha.  The dotted lines mark the edges of the echelle slit used in Figure~1.}
\end{figure}

\begin{figure}
\epsscale{.5}\plotone{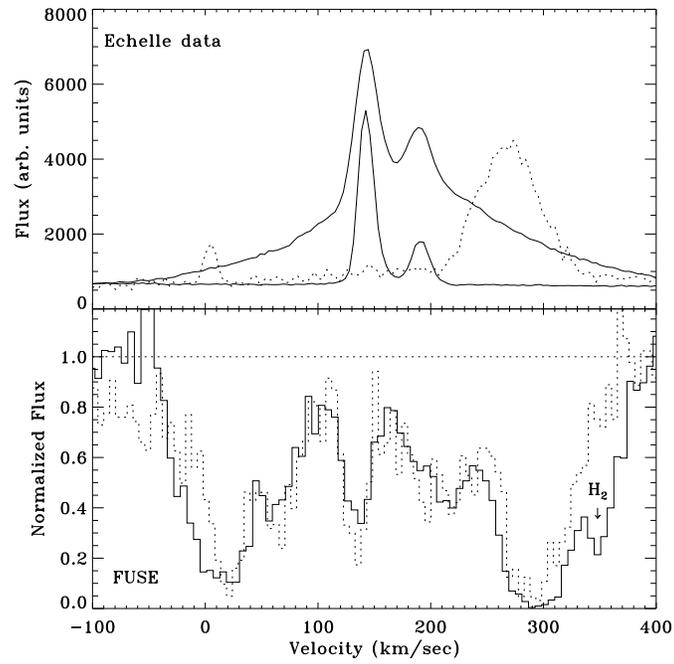}
\caption{Top panel: an extracted spectrum of the star in velocity space. The two solid lines show \Ha\ and [\NII] $\lambda$6583 (\Ha\ is the one with the broad stellar emission peak).  The dashed curve shows the spectrum of the ISM immediately to the east and west of S119. Note the zero velocity atmospheric \Ha\ line.  Bottom panel: two normalized FUSE spectra;  \OI\ $\lambda$1039 (solid) and \FeII\ $\lambda$1145 (dashed).  Five distinct absorption features can be seen in each profile data.  The feature at 350 \kms\ in \OI\ is \HH\ $\lambda$1040.37 at Galactic velocities.}
\end{figure}


\begin{thebibliography}{S119}
\bibitem[Blaauw 1961]{Blaauw61}
Blaauw, A. 1961, Bull. Astron. Inst. Netherlands, 15, 265
\bibitem[Danforth et al. in prep]{Danforth02}
Danforth, C. W. et al. in prep
\bibitem[Moos et al. (2000)]{Moos00}
Moos, H. W., et al. 2000, \apj, 538, L1
\bibitem[Nota et al. 1994]{Nota94}
Nota, A., et al. 1994, in `Circumstellar Media in the Late Stages of Stellar Evolution', eds. R. Clegg et al. (Cambridge: Cambridge University Press), p.89
\bibitem[Nota et al. 1995]{Nota95}
Nota, A., Livio, M., Clampin, M., \& Schulte-Ladbeck, R. 1995, \apj, 448, 788
\bibitem[Richter et al. 1999]{Richter99}
Richter, P., et al. 1999, \nat, 402, 386
\bibitem[Roth \& Blades (1997)]{RothBlades97}
Roth, K. C., \& Blades, J. C. 1997 \apj, 474, L95
\bibitem[Sahnow et al. (2000)]{Sahnow00}
Sahnow, D. J., et al. 2000, ApJ, 538, L7
\bibitem[Smith et al. 1998]{Smith98}
Smith, L., et al. 1998, \apj, 503, 278
\bibitem[Tweedy \& Kwitter 1996]{TK96} 
Tweedy, R. W., \& Kwitter, K. B. 1996, \apjs, 107, 255
\bibitem[Walborn 1982]{Walborn82}
Walborn, N. 1982, \apj, 256, 452
\end{thebibliography}
\end{document}